{}
{}

\documentclass[prd,reprint,superscriptaddress,preprintnumbers]{revtex4-1}

\newif\ifpublic\publictrue

\usepackage{fancyhdr}
\ifpublic\else
\fi
\newif\ifworking\workingtrue

\usepackage{mathtools,mathrsfs,amsbsy,amssymb,latexsym,amsfonts,amscd,
	amsmath,, multirow}
\usepackage{bm}

\usepackage{graphicx}
\usepackage[usenames,dvipsnames]{color}
\usepackage[normalem]{ulem}
\usepackage{natbib}
\usepackage{xcolor}
\usepackage{comment}
\definecolor{linkcolor}{rgb}{0,0,0.6}
\usepackage[ pdftex,colorlinks=true,
pdfstartview=FitV,
linkcolor= linkcolor,
citecolor= linkcolor,
urlcolor= linkcolor,
hyperindex=true,
hyperfigures=false]
{hyperref}

\newcommand{\SU}[1]{\mathrm{SU}( #1 )}

\newcommand{\SL}[1]{\mathrm{SL}( #1 )}

\newcommand{\U}[1]{\mathrm{U}(#1)}

\newcommand{\Sp}{\textrm{S}}

\def\showkeysrefformat#1{{\normalfont\tiny\ttfamily#1}}
\makeatletter
\def\SK@@ref#1>#2\SK@{%
	{\@inlabelfalse\leavevmode\vbox to\z@{%
			\vss\SK@refcolor\rlap{\vrule\raise .75em%
				\hbox{\showkeysrefformat{#2}}}}}}
\makeatother

\allowdisplaybreaks[3]

\begin{document}
\title{Holographic Evidence for Non-Supersymmetric Conformal Manifolds}

\author{Alfredo Giambrone}
\email{alfredo.giambrone@polito.it}
\affiliation{Department of Applied Science and Technology, Politecnico di Torino,
Corso Duca degli Abruzzi 24, I-10129 Torino, Italy; INFN--Sezione di Torino,
Via P. Giuria 1, 10125 Torino, Italy}

\author{Adolfo Guarino}
\email{adolfo.guarino@uniovi.es}
\affiliation{Departamento de F\'isica, Universidad de Oviedo, Avda. Federico Garc\'ia Lorca 18, 33007 Oviedo, Spain}
\affiliation{Instituto Universitario de Ciencias y Tecnolog\'ias Espaciales de Asturias (ICTEA), Calle de la Independencia 13, 33004 Oviedo, Spain}

\author{Emanuel Malek}
\email{emanuel.malek@physik.hu-berlin.de}
\affiliation{Institut f\"ur Physik, Humboldt-Universit\"at zu Berlin, IRIS Geb\"aude, Zum Gro{\ss}en Windkanal 2, 12489 Berlin, Germany}

\author{Henning Samtleben}
\email{henning.samtleben@ens-lyon.fr}
\affiliation{Univ Lyon, Ens de Lyon, CNRS,	Laboratoire de Physique, F-69342 Lyon, France}
\affiliation{Institut Universitaire de France (IUF), France}

\author{Colin Sterckx}
\email{colin.sterckx@ulb.ac.be}
\affiliation{Universit\'e Libre de Bruxelles (ULB) and International Solvay Institutes,
Service de Physique Th\'eorique et Math\'ematique,
Campus de la Plaine, CP 231, B-1050, Brussels, Belgium}
\affiliation{Departamento de F\'isica, Universidad de Oviedo, Avda. Federico Garc\'ia Lorca 18, 33007 Oviedo, Spain}

\author{Mario Trigiante}
\email{mario.trigiante@polito.it}
\affiliation{Department of Applied Science and Technology, Politecnico di Torino,
Corso Duca degli Abruzzi 24, I-10129 Torino, Italy; INFN--Sezione di Torino,
Via P. Giuria 1, 10125 Torino, Italy}

\preprint{HU-EP-21/58}

\begin{abstract}
We provide the first holographic evidence for the existence of a non-supersymmetric conformal manifold arising from exactly marginal but supersymmetry-breaking deformations of a superconformal three-dimensional field theory. In particular, we construct a 2-parameter non-supersymmetric deformation of a supersymmetric AdS non-geometric background in Type IIB string theory. We prove that the non-supersymmetric backgrounds are perturbatively stable and also do not suffer from various non-perturbative instabilities. Finally, we argue that diffeomorphism symmetry protects our solutions against higher-derivative string corrections.

\end{abstract}

\maketitle
Amongst quantum field theories, conformal field theories (CFTs) play a distinguished role. For example, CFTs are important in statistical mechanics, where they provide a description of many phase transitions. Moreover, CFTs are fixed points of the renormalisation-group flow thus introducing a notion of universality. Finally, because of the constraints imposed by conformal invariance, strongly-coupled CFTs can provide an insight into non-perturbative QFTs more generally.

An important question to ask when studying CFTs is whether they are isolated fixed points of the renormalisation-group flow, or belong to a family of CFTs, known as a conformal manifold. The conformal manifold is spanned by exactly marginal deformations of the CFT, i.e.\ marginal operators whose $\beta$-functions vanish exactly to all orders. Over the last decade, much insight has been gained into local properties of conformal manifolds of supersymmetric conformal field theories \cite{Leigh:1995ep,Kol:2002zt,Green:2010da,Kol:2010ub,Baggio:2017mas}. In particular, it is not uncommon for four-dimensional ${\cal N}=1$ and three-dimensional ${\cal N}=2$ CFTs to possess conformal manifolds, whose dimensions can be deduced from the symmetry of the CFTs, without need to compute $\beta$-functions or even having a Lagrangian description.

On the other hand, no example is known  of a  non-supersymmetric conformal field theory in more than two dimensions featuring a conformal manifold. Indeed, they are widely believed not to exist, since it is unclear how the precise cancellations in the $\beta$-functions will be achieved without supersymmetry. However, there are no ``no-go theorems'' that forbid non-supersymmetric conformal manifolds. As a result, the existence of non-supersymmetric conformal manifolds has been largely the subject of speculation, with only few systematic analyses performed recently \cite{Bashmakov:2017rko,Behan:2017mwi,Hollands:2017chb,Sen:2017gfr}.

The AdS/CFT correspondence \cite{Maldacena:1997re,Gubser:1998bc,Witten:1998qj} between anti-de Sitter (AdS) solutions of string theory and CFTs provides a powerful tool to address this question, at least in the ``large-N limit'' where the rank of the gauge group of the CFT is taken to be large. The correspondence maps the conformal manifold of a CFT to a continuous family, known as the ``moduli space'', of AdS solutions of string theory. As yet, no continuous family of non-supersymmetric AdS solutions of string theory has been constructed, with the possible exception of \cite{Frolov:2005dj} (see discussion below). Indeed, non-supersymmetric AdS solutions of string theory are conjectured to be unstable \cite{Ooguri:2016pdq}, with only a handful of \textit{isolated} potentially stable non-supersymmetric AdS vacua known \cite{Guarino:2020flh}.

In this letter, we will provide the first holographic evidence for a three-dimensional non-supersymmetric conformal manifold. We do this by constructing a 2-parameter non-supersymmetric deformation of an ${\cal N}=4$ supersymmetric AdS$_4$ vacuum describing a non-geometric solution of Type IIB superstring theory. We will prove that the entire 2-parameter family is perturbatively stable in IIB supergravity, and show that it does not suffer from various non-perturbative instabilities. We note that just as for the supersymmetric deformations considered in \cite{Giambrone:2021zvp,Guarino:2021kyp}, the non-supersymmetric deformations we study here can also locally be absorbed by coordinate redefinitions, which are, however, not globally well-defined. This implies that any local diffeomorphism-invariant quantities, such as those controlling higher-derivative corrections of string theory, are independent of the deformations. This provides hope that our conformal manifold may also exist beyond the large-N limit of the CFT.

We construct our non-supersymmetric 2-parameter family of AdS$_4$ vacua of IIB string theory by uplifting the corresponding family of AdS$_4$ vacua of four-dimensional $[{\rm SO}(6)\times {\rm SO}(1,1)]\ltimes \mathbb{R}^{12}$ supergravity \cite{Guarino:2021hrc} using the truncation Ansatz of \cite{Inverso:2016eet}. Our family of AdS$_4$ vacua depends on two ``axionic'' parameters $\chi_{1}$, $\chi_2$ \cite{Guarino:2021hrc}. For generic values of $\chi_{1,2}$, the AdS$_4$ vacua are non-supersymmetric and preserve a $\textrm{U}(1)^2$ symmetry. Three patterns of (super) symmetry enhancement occur at special loci of the $(\chi_{1},\chi_{2})$ parameter space. For $\chi_1 = \pm \chi_2$, there is an ${\cal N}=2$ supersymmetry enhancement whereas a $\textrm{U}(1)^2$ symmetry is still preserved. These AdS$_4$ vacua belong to the family considered in \cite{Guarino:2020gfe,Giambrone:2021zvp,Bobev:2021yya}. For $\chi_{1}=0$ or $\chi_{2}=0$, the vacua are non-supersymmetric but the residual symmetry gets enhanced to $\textrm{SU}(2) \times \textrm{U}(1)$. Lastly, an $\mathcal{N}=4$ and SO(4) symmetric AdS$_{4}$ vacuum appears at the special point $\chi_{1}=\chi_{2}=0$. As a result, $\chi_{1,2}$ parameterise non-supersymmetric deformations of the ${\cal N}=4$ AdS$_4$ S-fold vacuum of IIB string theory \cite{Inverso:2016eet}.

The ten-dimensional geometry we obtain is a non-supersymmetric ``S-fold'' of the form $\textrm{AdS}_{4} \times \Sp_{\eta}^1 \times \Sp^5$, where $\Sp^5 = \mathcal{I} \times \Sp_{1}^2 \times \Sp_{2}^2$ and $\mathcal{I} $ is an interval with angular coordinate $\alpha \in [0,\tfrac{\pi}{2}]$. The term S-fold refers to the fact that the 10-dimensional solution has an $\SL{2,\mathbb{Z}}$ S-duality monodromy of IIB string theory as we move around the $\Sp^1_\eta$ circle. The corresponding dual CFT is known as a J-fold CFT obtained by compactifying ${\cal N}=4$ super Yang-Mills theory on a circle with an $\SL{2,\mathbb{Z}}$ twist \cite{Assel:2018vtq}. Holography has recently proven powerful in studying supersymmetric AdS$_4$ vacua of these types and their supersymmetric deformations \cite{Inverso:2016eet,Arav:2021tpk,Arav:2021gra,Giambrone:2021zvp,Guarino:2021hrc,Bobev:2021rtg,Guarino:2021kyp,Bobev:2021yya,Cesaro:2021tna,Berman:2021ynm}.

More concretely, the S-fold solution can be constructed out of the following 10-dimensional solution of classical Type IIB supergravity:
\begin{equation}
\label{metric_10D_solu}
\begin{split}
ds_{10}^2 &= \Delta^{-1}  \left[\tfrac{1}{2} \, ds_{\textrm{AdS}_4}^2+ d\eta^2 + d\alpha^2 \right. \\
 &\quad + \left.\dfrac{\cos^2\alpha}{2+\cos(2\alpha)} d\Omega_{1}  + \dfrac{\sin^2\alpha}{2-\cos(2\alpha)} d\Omega_{2} \right] \,,
\end{split}
\end{equation}
where $\chi_{i}$-twisted two-spheres $\Omega_{i}$ have metrics
\begin{equation}
d\Omega_{i} = d\theta_{i}^2 + \sin^2\theta_{i} \, {d\varphi'_{i}}^2 
\hspace{3mm} \textrm{ with } \hspace{3mm}
d\varphi'_{i} = d\varphi_{i} + \chi_{i} \, d\eta \ ,\label{phieta}
\end{equation}
and the non-singular warping factor is
\begin{equation}
\Delta^{-4} = 4 - \cos^2(2\alpha) \ . 
\end{equation}
The two-form potential $B_{2}$ and $C_{2}$ take the form
\begin{equation}
\label{B2-C2_alt}
\begin{split}
B_{2} &= - 2 \sqrt{2} \, e^{-\eta} \, \dfrac{\cos^3\alpha}{2 + \cos(2\alpha)} \, \textrm{vol}_{\Omega_1} \,,\\
C_{2} &= - 2 \sqrt{2} \, e^{\eta} \, \dfrac{\sin^3\alpha}{2 - \cos(2\alpha)} \, \textrm{vol}_{\Omega_2} \,,
\end{split}
\end{equation}
whereas the dilaton $g_{s}=e^{\Phi}$ and the axion $C_{0}$ read
\begin{equation}
e^{\Phi} =\sqrt{2} \,  e^{-2 \eta} \, \dfrac{2-\cos(2\alpha)}{\sqrt{7- \cos(4\alpha)}}\,, 
\hspace{5mm} \textrm{ and } \hspace{5mm} 
C_{0} = 0 \,.
\end{equation}
The four-form potential $C_{4}$, yielding a self-dual field strength $\widetilde{F}_{5} = dC_{4} + \frac{1}{2}  \left(  B_{2} \wedge dC_{2}  - C_{2} \wedge dB_{2}\right)$, reads
\begin{equation}
\begin{split}
\label{C4SD_potential}
C_{4} &= \tfrac{3}{2} \, \omega_{3} \wedge  \left( d\eta + \tfrac{2}{3} \sin{(2 \alpha)} \, d\alpha \right) \\
& \quad - \tfrac{1}{2} \, f(\alpha)  \,  d\alpha  \wedge \left( A_{1} \wedge \textrm{vol}_{\Omega_2} + \textrm{vol}_{\Omega_1} \wedge A_{2}  \right)  \,,
\end{split}
\end{equation}
where $d\omega_{3}=\textrm{vol}_{\textrm{AdS}_{4}}$ with AdS radius $L_{\textrm{AdS}_{4}}=1$. The function $f(\alpha)$ in (\ref{C4SD_potential}) is given by
\begin{equation}
\label{f_func}
f(\alpha) = \sin ^2(2 \alpha )  \, \frac{\cos (4 \alpha)-55}{\big(7-\cos (4 \alpha) \big)^2} \ ,
\end{equation}
where we have introduced one-forms $A_i = - \cos\theta_{i} \, d\varphi'_{i}$ so that $dA_i = \textrm{vol}_{\Omega_i}$. Note that the function $f(\alpha)$ in (\ref{f_func}) vanishes at $\alpha=0,\,\frac{\pi}{2}$, where each of the $\Sp^2$ shrinks to zero size in a smooth way so that the compact space is topologically $\Sp^1_\eta \times \Sp^5$. We have explicitly verified that the above class of backgrounds satisfies the ten-dimensional Type IIB equations of motions and source-less Bianchi identities.

The S-fold solution, characterised by an ${\rm SL}(2,\mathbb{Z})$ monodromy along $\Sp^1_\eta$, can then be obtained from the above solution through a suitable ${\rm SL}(2,\mathbb{R})$-transformation together with an appropriate choice of the period $T$, according to the prescription given in \cite{Inverso:2016eet,Assel:2018vtq}. In this way the monodromy can be chosen, for instance, to be a hyperbolic element of the form
\begin{equation}
    J_k = \begin{pmatrix}
     k & 1 \\ -1 & 0
    \end{pmatrix} \,, \qquad k > 2 \,.
\end{equation}
This choice requires the $\Sp^1_\eta$ radius to be
\begin{equation}
    T = \log\left(k+\sqrt{k^2-4}\right) - \log{2}\,.
\end{equation}
Moreover, $k$ can be chosen such that the supergravity approximation remains valid throughout the S-fold solution, because the dilaton and derivatives of the axio-dilaton remain small throughout \cite{Assel:2018vtq}. We shall refrain from further discussing these aspects of the solution, since they do not affect our present analysis, which focuses on the 2-parameter deformation of the background and is independent of the duality twist.

The $\chi_{1,2}$ deformations only appear in the background via the combination \eqref{phieta} and thus can locally be absorbed by the coordinate redefinition
\begin{equation} \label{eq:LocalDiffeoDelta1}
    \varphi'_i = \varphi_i + \chi_i\, \eta \,.
\end{equation}
However, due to the periodicity of $\eta \rightarrow \eta + T$, this is only a well-defined coordinate transformation when $\chi_i = \frac{2\pi k_i}{T}$ for $k_i \in \mathbb{Z}$. This suggests that the deformation parameters are periodic with period $\frac{2\pi}{T}$. However, there is a subtlety because of how the spinors are defined on the $\Sp^1_\eta$. In fact, by looking at the spinors, as we will demonstrate later in \eqref{eq:SpectrumShift} through the Kaluza-Klein spectrum, we see that the correct periodicity is in fact $\chi_i \in \left[0,\frac{4\pi}{T}\right)$. This means that the non-supersymmetric conformal manifold is compact and has topology $T^2/\mathbb{Z}_2$, where the $\mathbb{Z}_2$ corresponds to the interchange $\chi_1 \leftrightarrow \chi_2$.

An alternative description of the parameters $\chi_i$ comes from their oxidation to the five-dimensional supergravity obtained by reducing IIB string theory on $\Sp^5$. As noted in \cite{Guarino:2021kyp,Guarino:2021hrc} (see also \cite{Berman:2021ynm,Bobev:2021rtg}) the $\chi_i$ define non-trivial one-form deformations (Wilson loops) for the vector fields along $\Sp^1_\eta$. For the $\mathcal{N}=4$ S-fold, this corresponds to turning on Wilson loops for the $\SU{2} \times \SU{2}$-valued gauge fields breaking the symmetry down to its Cartan subgroup.

It is instructive to compare the deformation of the $\mathcal{N}=4$ S-fold solution analysed here, with the deformation, discussed in \cite{Frolov:2005dj}, of the maximally supersymmetric ${\rm AdS}_5\times \Sp^5$ Type IIB background, which generalises the Lunin-Maldacena construction \cite{Lunin:2005jy}. The holographic dual to this solution is conjectured to be  a non-supersymmetric marginal  deformation of $\mathcal{N}=4$ four-dimensional SYM theory. However, \cite{Fokken:2013aea} suggested that conformal symmetry of this dual theory is absent, while \cite{Russo:2005yu,Spradlin:2005sv} hint at the existence of a tachyonic instability in the corresponding superstring background. In \cite{Frolov:2005dj}, the deformation parameters $\gamma_I$, $I=1,2,3$, were the effect of shift transformations in the ${\rm O}(3,3)$ group acting on the three angular directions associated with translational isometries \cite{Giveon:1991jj} along internal angular coordinates. These shift transformations were, however, preceded and followed by T-dualities  (``factorised dualities'') of the kind $R_I\rightarrow 1/R_I$ along all the three directions.  Just as $\Sp^5$ in the ${\rm AdS}_5\times \Sp^5$ background, the internal manifold $\mathcal{I}\times \Sp_1^2\times \Sp_2^2\times \Sp^1_\eta$ of the $\mathcal{N}=4$ S-fold solution features three angular coordinates $\xi^I=\varphi_1,\,\varphi_2,\,\eta$ each associated with a translational symmetry of the internal metric, although, in the latter case, a constant translation along $\eta$ is not a symmetry of the whole solution due to the ${\rm SL}(2,\mathbb{R})_{{\rm IIB}}$-twist. As opposed to the construction of \cite{Frolov:2005dj}, the $\chi_i$-deformation discussed here only results from a shift transformation in ${\rm GL}(3,\mathbb{R})\subset {\rm O}(3,3)$, with no T-dualities. This is effected by 
 the ${\rm GL}(3,\mathbb{R})$ matrix
 \begin{equation}
A=\left(\begin{matrix}1 & 0 & \chi_1\cr 0 & 1 & \chi_2\cr 0 & 0 & 1\end{matrix}\right)\,,\label{Amat}
\end{equation}
which acts linearly on the $I$-component of all the fields. The components $g=(g_{IJ})$ of the internal metric along the angular directions $\xi^I$, for instance, transforms as follows:
\begin{equation}
g\rightarrow A^t\,g\,A\,.
\end{equation}
Our $\chi_i$ deformations thus change the metric on the $\Sp^5 \times \Sp^1_\eta$ compactification, while leaving the fibration structure unchanged. This is analogous to complex structure deformations of $T^2 \sim \Sp^1 \times \Sp^1$, which can also locally be absorbed by diffeomorphisms which are, however, not globally well-defined. Indeed, our $\chi_i$ appear like the real part of complex structure deformations of the $\varphi_i \times \Sp^1_\eta$ tori. A more precise definition is in terms of the mapping torus of $\Sp^5$ \cite{Guarino:2021hrc}: the $\chi_i$ deformations imply that as we move around $\Sp^1_\eta$, we end up in a different point on $\Sp^5$. If $\chi_i\rightarrow \chi_i+2\pi k_i/T$, $k_i\in \mathbb{Z}$, the deformation is in ${\rm GL}(3,\mathbb{Z})$ and the internal geometry is not affected. Invariance of the full spectrum, however, due to the presence of states with half-integer $j_1,\,j_2$, extends the periodicity of $\chi_i$ to $4\pi/T$, as will be discussed below.

Via the AdS/CFT correspondence, our family of non-supersymmetric AdS$_4$ vacua of IIB string theory suggests that the dual ``J-fold'' CFT$_3$ should belong to a non-supersymmetric conformal manifold. However, this is not the case if the non-supersymmetric AdS$_4$ vacua are unstable, as conjectured in \cite{Ooguri:2016pdq}. These instabilities could arise due to some scalar fluctuation in the Kaluza-Klein spectrum violating the Breitenlohner-Freedman bound, or via a non-perturbative phenomenon. Let us now address these concerns.

First, we will prove that the Kaluza-Klein spectrum has no tachyons, i.e.\ the AdS$_4$ vacua are perturbatively stable. To do this, we use the technology developed in \cite{Malek:2019eaz,Malek:2020yue} to compute the full Kaluza-Klein spectrum around the family of non-supersymmetric AdS$_4$ vacua we consider here.

It is easiest to express the Kaluza-Klein spectrum as a deformation of the spectrum of the ${\cal N}=4$ vacuum. The full ${\cal N}=4$ spectrum was computed in \cite{Dimmitt:2019qla,Giambrone:2021zvp}. Note that our $\Sp^1$ radius differs from the convention of \cite{Giambrone:2021zvp} such that $T_{\rm there} = \frac{T_{\rm here}}{2}$. The conformal dimension of the highest weight state of each supermultiplet is given by
\small
\begin{equation} \label{eq:SpectrumN4}
 \Delta = \frac32 + \frac12 \sqrt{9 + 2\ell(\ell+4) + 4 \sum_{i=1,2}\ell_i(\ell_i+1)+ 2\left(\frac{2 n \pi}{T}\right)^2} \,,
\end{equation}
\normalsize
where $\ell$ denotes the $\Sp^5$ Kaluza-Klein level, $n$ the $\Sp^1$ level and $\ell_1$, $\ell_2$ the $\textrm{SO}(4)$ spin of the highest weight state (in this case, the graviton). These ${\cal N}=4$ supermultiplets are counted by the generating function for their highest weight states:
\begin{equation} \label{eq:Character}
    \nu = \frac{1}{(1-q^2)(1-q\,u)(1-q\,v)} \frac{1+s}{1-s} \ ,
\end{equation}
where the exponent of $q$ and $s$ determine the Kaluza-Klein levels on the $\Sp^5$, $\ell$, and $\Sp^1$, $n$, while the exponents of $u$ and $v$ count the $\SU{2} \times \SU{2}$ spins, $\ell_1$ and $\ell_2$. The effect of the $\chi_{1,2}$ deformations is to shift the conformal dimension of each field by replacing
\begin{equation} \label{eq:SpectrumShift}
 \frac{2n\pi}{T} \longrightarrow 
 \frac{2n\pi}{T} + (j_1+j_2) \chi_+ + (j_1- j_2) \chi_- \,,
\end{equation}
in \eqref{eq:SpectrumN4}, where $j_1$, $j_2$ are the charges of the field under the $\U{1} \times \U{1}$ Cartan of $\textrm{SO}(4)$ and we defined $\chi_\pm = \tfrac12 (\chi_1 \pm \chi_2)$. Note from \eqref{eq:Character} that, while $j_1$, $j_2$ are half-integers, $j_1 \pm j_2$ are always integers. Thus, we manifestly see that the full background has periodicity $\chi_\pm \rightarrow \chi_\pm + \frac{2\pi}{T}$, upon which the Kaluza-Klein spectrum gets mapped back to itself with an appropriate reshuffling of the fields amongst the $\Sp^1$ level with $n \rightarrow n - (j_1\pm j_2)$, just like in \cite{Giambrone:2021zvp}. Notice that $\chi_1,\,\chi_2$ separately have period $4\pi/T$, which can only be seen from the spinors on the AdS$_4$ background which have half-integers charges under the $\U{1} \times \U{1}$ Cartan.

Even more importantly, we can see that the masses for all the fields are bounded from below by the masses of the fields of the four-dimensional supergravity at the ${\cal N}=4$ vacuum, i.e.\ where $\ell = \ell_1 = \ell_2 = n = \chi_i = 0$. This in particular implies that all scalars have masses above the Breitenlohner-Freedman bound for any value of $\chi_i$. Thus, the non-supersymmetric vacua are perturbatively stable.

One may also wonder whether the AdS$_4$ vacua are secretly supersymmetric in 10 dimensions, with some gravitinos amongst the higher Kaluza-Klein modes becoming light, akin to the ``space invaders'' scenario \cite{Duff:1986hr,Cesaro:2020piw,Giambrone:2021zvp}. However, from \eqref{eq:SpectrumN4}, \eqref{eq:SpectrumShift}, we can easily see that such gravitinos can only restore supersymmetry when the combination $\frac{2n\pi}{T} + j_1 \chi_1 + j_2 \chi_2 = 0$. This can only occur when either $n = 0$ and $\chi_1 = \pm \chi_2$, corresponding to supersymmetry enhancement that already occurs in the four-dimensional supergravity \cite{Guarino:2021hrc}, or $\chi_\pm = \frac{2\pi k_\pm}{T}$, for $k_\pm \in \mathbb{Z}$ when some gravitinos at $\Sp^1$ level $n = - (j_1+j_2) k_+ - (j_1-j_2) k_-$ become massless. This latter condition is precisely when the background is mapped back to itself, so that for $0 < \chi_\pm < \frac{2\pi}{T}$, $\chi_1 \neq \pm \chi_2$, the AdS$_4$ vacua are not supersymmetric in the full Type IIB string theory.

Next we investigate the non-perturbative stability of the non-supersymmetric AdS$_4$ vacua. Since the AdS$_4$ vacua arise as near-horizon limits of certain brane configurations, one may worry that for the non-supersymmetric vacua the corresponding brane configurations become unstable \cite{Maldacena:1998uz}. We search for signs of such instabilities by considering single probe D$p$-branes (and single probe NS5-branes) with rigid embeddings in our AdS$_4$ vacua. In particular, we check whether the branes are unstable due to a greater repulsive force of the fluxes coming from the WZ term than the attractive (i.e.\ towards the interior of the AdS spacetime) gravitational force due to the DBI term. Indeed, \cite{Ooguri:2016pdq} conjectures that there should always be some branes that are unstable in this way, see also \cite{Bena:2020xxb}. However, we find that single probe D$p$-branes and NS5-branes without worldvolume flux remain stable when placed in the non-supersymmetric backgrounds (\ref{metric_10D_solu})--(\ref{f_func}).

The stability of these probe branes can be understood in the following way. Firstly, note that we can perform the diffeomorphism \eqref{eq:LocalDiffeoDelta1} to remove the $\chi_i$ deformation from the metric. However, now the coordinates respect the deformed periodicities
\begin{equation}
    \begin{split}
        \varphi'_i &\rightarrow \varphi'_i + 2\pi \,, \\
        \eta &\rightarrow \eta + T \,, \quad \varphi'_i \rightarrow \varphi_i' + \chi_i\, T \,.
    \end{split}
\end{equation}
As a result, the only well-defined embeddings of branes wrapping $\eta$ must also wrap $\varphi'_i$. In particular, let us denote by $\xi  \sim \xi + T$ the relevant wrapped worldvolume coordinate on the brane. Then, the only well-defined embeddings are given by
\begin{equation}
    \begin{split}
        \eta(\xi)=q\,\xi \,, \qquad \varphi'_{i}(\xi)=\left(p_{i} \frac{2 \pi}{qT} + \chi_{i}\right)\xi \,,
    \end{split}
\end{equation}
with $p_{i} \in \mathbb{Z}$. We see that as $\chi_i$ is turned on, a brane wrapping $\Sp^1_\eta$ must also wrap increasing amounts of $\varphi'_i$, so that the DBI part of the action increases. At the same time, for $p$-branes, with $p \neq 5$, the WZ coupling is insensitive to wrapping along $\varphi'_i$, unless the brane is completely internal. Therefore, these branes either become more stable as $\chi_i$ are turned on or they are completely internal branes, which cannot trigger non-perturbative instabilities in the usual way. Finally, an explicit computation for NS5- and D5-branes shows that they also remain stable as $\chi_i$ are turned on in the backgrounds \eqref{metric_10D_solu} -- \eqref{f_func}. 



Finally, non-supersymmetric vacua may also decay due to bubbles of nothing \cite{Witten:1981gj}, which requires a circle or sphere \cite{Ooguri:2017njy} to collapse. However, our internal space $\Sp^5 \times \Sp_\eta^1$ is topologically protected from such a collapse: the $\Sp^5$ cannot collapse as it is supported by flux, whereas the $\Sp_\eta^1$ cannot collapse since the spinors do not have anti-periodic boundary conditions on it \cite{Witten:1981gj}, but instead general periodicities along $\Sp^1_\eta$, provided $(\chi_1,\chi_2)\neq (\frac{2\pi}{T},0), \,(0,\frac{2\pi}{T})$. This means that a straightforward bubble of nothing cannot occur. Still, our vacua could decay semi-classically via more complicated bubbles of nothing containing defects, e.g. a D3-brane in $\Sp^5$ similar to \cite{Horowitz:2007pr,Bomans:2021ara} or an O7-plane in $\Sp^1$ \cite{McNamara:2019rup}. However, because the volume form of the compactification is independent of the $\chi_i$ deformations, our non-supersymmetric AdS$_4$ vacua are likely to be stable against the instanton decay constructed in \cite{Bomans:2021ara}, which is \emph{delocalised} on the compactification space. On the other hand, constructing the \emph{localised} instanton solutions is extremely technically challenging. Moreover, the mechanism of \cite{Bomans:2021ara} treats a shrinking dilaton as equivalent to a shrinking $\Sp^1$. Aside from the validity of this equivalence, a similar shrinking dilaton would be problematic for our S-fold vacua, where the dilaton is not well-defined due to the $\SL{2,\mathbb{Z}}$ monodromy along $\Sp^1_\eta$.

So far, we have proven that our AdS$_4$ vacua are perturbatively stable and have provided evidence that they may also be stable against semi-classical decay. However, one may worry that while our AdS$_4$ geometries are solutions of IIB supergravity, the higher-derivative corrections of IIB string theory will spoil our solutions. In the dual CFT, this would imply that some $\frac{1}{N}$ corrections lift the conformal manifold. However, the deformations $\chi_i$ can always be locally absorbed by the coordinate redefinition \eqref{eq:LocalDiffeoDelta1}, which however is not globally well-defined. Therefore, all local diffeomorphism-invariant quantities are independent of the $\chi_i$. In particular, this means that each term of the higher-derivative corrections of string theory, involving powers of the curvature tensor or the fluxes, are also independent of $\chi_{1,2}$. Thus, our non-supersymmetric AdS$_4$ vacua are equally valid solutions of IIB string theory as the ${\cal N}=4$ vacuum. Moreover, the $\chi_{i}$ deformations actually correspond to parity-odd (pseudo) scalars in the maximal supergravity \cite{Guarino:2021hrc}, so the potential $1/N$ tadpole destabilisation of \cite{Berkooz:1998qp} cannot take place for our backgrounds.

There could still be some string corrections, e.g. from branes wrapping the compactification, which are sensitive to $\chi_{i}$ and which could thus spoil our solutions. For example, D$p$-instantons could wrap some $(p+1)$-cycle of the compactification, and depend on $\chi_{i}$. However, our solutions are also protected against such instanton corrections, since the compactification $\Sp^5 \times \Sp^1_\eta$ only has non-trivial 1-, 5- and 6-cycles. Therefore, we can only have D5-instantons wrapped on the full $\Sp^5 \times \Sp^1_\eta$. But since the volume form is independent of $\chi_{i}$, these instantons give no corrections to our solutions. Nonetheless, one could expect some other extended state to do so, corresponding to some $\frac1N$ correction in the dual CFT.

According to the proposal put forward in \cite{Assel:2018vtq}, the SCFT dual to the $\mathcal{N}=4$ background emerges as the effective IR description of a $3d$ ${\rm T}[{\rm U}(N)]$ theory \cite{Gaiotto:2008ak} in which the diagonal subgroup of the ${\rm U}(N)\times {\rm U}(N)$ flavour group has been gauged using an $\mathcal{N}=4$ vector multiplet. In addition, a Chern-Simons term at level $k$ must be introduced which is defined by the $J_k=-\mathcal{S}\,\mathcal{T}^k\in {\rm SL}(2,\mathbb{Z})_{{\rm IIB}}$ monodromy along the $\Sp^1_\eta$. The effective $\mathcal{N}=4$ superpotential \cite{Beratto:2020qyk} $W_{\textrm{eff}} = (2\pi/k) \, {\rm Tr}(\mu_H\,\mu_C)$ is marginal in the IR and, in \cite{Bobev:2021yya}, a shift $W_{\textrm{eff}} \rightarrow W_{\textrm{eff}} \, + \, \lambda \, {\rm Tr}(\mu_H\,\mu_C)$ with $\lambda \in \mathbb{C}$ was proposed as an exactly marginal deformation preserving $\mathcal{N}=2$. The scalar superconformal primary operators $\mu_H$ and $\mu_C$ are respectively described by the moment maps of the Higgs and Coulomb branch of ${\rm T}[{\rm U}(N)]$. Each of the $\mu_H$ and $\mu_C$ fields is a component of a vector in the adjoint representation of the corresponding ${\rm SU}(2)$ subgroup of the ${\rm SO}(4)$ R-symmetry group (to be denoted by ${\rm SU}(2)_H$ and ${\rm SU}(2)_C$, respectively). We can therefore associate with $\mu_H$ the quantum numbers $j_1=1,\,j_2=0$ and with $\mu_C$ the values $j_1=0,\,j_2=1$, having identified $j_1,\,j_2$ with the eigenvalues of the Cartan generators of ${\rm SU}(2)_H$ and ${\rm SU}(2)_C$, respectively. Note that  $\chi_1$ ($\chi_2$) only breaks ${\rm SU}(2)_H$ (${\rm SU}(2)_C$) to its ${\rm U}(1)_H$ (${\rm U}(1)_C$) subgroup. The combination $(\chi_1-\chi_2)/2$ of these two parameters, for $\chi_1=-\chi_2$, should already be encoded in the $\lambda$ parameter of the $\mathcal{N}=2$ exactly marginal deformation proposed in \cite{Bobev:2021yya}. We suggest that the orthogonal combination $(\chi_1+\chi_2)/2$,  be encoded in the conjectured exactly marginal deformation of the $3d$ Lagrangian:
\begin{equation}
\partial_\alpha \mathcal{O} \, \partial^\alpha \bar{\mathcal{O}} \ ,\label{OOb}
\end{equation}
where $\mathcal{O}\equiv  {\rm Tr}(\mu_H\,\bar{\mu}_C)$ is an operator with $j_1=1,\,j_2=-1$ and $\partial_\alpha$ denote the partial derivatives with respect to the (real) scalar fields. As opposed to ${\rm Tr}(\mu_H\,\mu_C)$, the above term does not originate from a holomorphic deformation of the superpotential and thus would  break all supersymmetries. The exact marginality of the operator (\ref{OOb}) is here conjectured in light of the holographic evidence we put forward.  Note that the resulting $\mathcal{N}=0$ theory would be parity symmetric in both the Higgs and the Coulomb sector: By performing, for instance, a parity transformation in the Coulomb sector which changes sign to the complex structure of the hyper-K\"ahler manifold (described as a complex  K\"ahler space),  $\mu_C\rightarrow \bar{\mu}_C$, and $\mathcal{O}$ would be exchanged with the exactly  marginal operator ${\rm Tr}(\mu_H\,{\mu}_C)$ in the superpotential proposed in \cite{Bobev:2021yya}. The same transformation would correspond in the bulk to a parity $\varphi_2\rightarrow -\varphi_2$ in $\Sp^2_2$ and, correspondingly, to $\chi_2\rightarrow -\chi_2$. It is therefore the simultaneous presence of the deformations $\mathcal{O}$, $\bar{\mathcal{O}}$ and ${\rm Tr}(\mu_H\,{\mu}_C)$ in the Lagrangian which breaks supersymmetry. Also, the $\chi_{1} \leftrightarrow \chi_{2}$ symmetry of the supergravity backgrounds amounts to an exchange symmetry between the Higgs and Coulomb branches in the dual non-supersymmetric CFT's.

Finally, our computation of the Kaluza-Klein spectrum \eqref{eq:SpectrumN4}, \eqref{eq:SpectrumShift} reveals not only the $\frac{4\pi}{T}$ periodicity of the exactly marginal deformations parameterised by $\chi_i$. It also gives the anomalous dimensions of all operators of the CFT dual to supergravity modes along the non-supersymmetric conformal manifold.

In this letter, we provided the first holographic evidence for the existence of a non-supersymmetric conformal manifold. We did this by constructing a 2-parameter family of non-supersymmetric S-fold AdS$_4$ vacua of IIB string theory, and proving that they are perturbatively stable. Moreover, we excluded several potential non-perturbative instability mechanisms, and showed that our solutions are even protected against some higher-derivative corrections.

Our findings here can be generalised and applied to other settings. For example, 
an analogous non-supersymmetric 2-parameter family of S-fold AdS$_4$ vacua can be obtained by performing 
similar axionic deformations to the $\U{1}$ R-symmetry and $\SU{2}$ flavour symmetry of the 
${\cal N}=2$ $\SU{2} \times \U{1}$ AdS$_4$ S-fold vacuum of IIB string theory \cite{Guarino:2020gfe}. This moduli space has a one dimensional locus of $\mathcal{N}=0$ deformations of the $\mathcal{N}=2$ $\SU{2} \times \U{1}$ vacuum, also contains the supersymmetric deformation studied in \cite{Giambrone:2021zvp} and should be connected to our conformal manifold since there is an exactly marginal deformation, connecting the ${\cal N}=2$ and ${\cal N}=4$ vacua \cite{Bobev:2021yya}. We explicitly verified that this second 2-parameter family is also perturbatively stable and has the same protection against non-perturbative mechanisms as was shown by our brane-jet computation and topological arguments. Moreover, the axionic deformations can again be reabsorbed by local coordinate redefinitions that fail to be globally well-defined \cite{Guarino:2021hrc}, yielding the same space-invaders scenario as here which leads to a $T^2$ moduli space. This also protects this 2-parameter family of AdS$_4$ vacua against higher-derivative corrections. Moreover, this same argument can be applied to the recently-constructed moduli space of ${\cal N}=1$ CFT$_3$'s \cite{Bobev:2021rtg}, which would suggest that also this ${\cal N}=1$ moduli space is protected against some higher-derivative corrections of string theory. The methods laid out here should also apply to a related class of S-folds where $\Sp^5$ is replaced by a Sasaki-Einstein manifold.

The fate of our family of non-supersymmetric AdS$_4$ vacua deserves further investigation. The brane-web whose near-horizon limit corresponds to the AdS$_4$ vacua could still suffer from some other instability mechanism. For example, it could feature some tachyon in its fluctuation spectrum, see e.g. \cite{Apruzzi:2019ecr,Apruzzi:2021nle} for recent discussions. However, because we do not know the brane-web that would give rise to the AdS$_4$ vacua, it is currently unclear which probe branes to use for this computation. Still, the existence of a continuous limit to the $\chi_{i}=0$ supersymmetric case could help in taming such potential instabilities. Also, some non-perturbative string corrections could lift the moduli space. Finally, the CFT$_3$ interpretation of the $\chi_i$ deformations deserves further exploration. We leave these exciting questions for future work.
\\[-2mm]

\noindent\textbf{Acknowledgements}:  
We are grateful to V. Bashmakov, M. Bertolini, A. Faedo, J. Gauntlett, H. Raj and D. Waldram for helpful discussions.
AG is supported by the Spanish government grant PGC2018-096894-B-100 and by the Principado de Asturias through the grant FC-GRUPIN-IDI/2018/000174. EM is supported by the Deutsche Forschungsgemeinschaft (DFG, German Research Foundation) via the Emmy Noether program ``Exploring the landscape of string theory flux vacua using exceptional field theory'' (project number 426510644). CS is supported by IISN-Belgium (convention 4.4503.15) and is a Research Fellow of the F.R.S.-FNRS (Belgium).

\bibliographystyle{utphys}
\bibliography{Biblio}

\end{document}